\documentclass{PoS}
\pdfoutput=1 
\newif\ifprlout

\usepackage{amssymb,amsmath}
\usepackage{graphicx}
\usepackage{dcolumn}
\usepackage{bm}

\newcommand{\ifb}{\ensuremath{\mathrm{fb^{-1}}}}
\newcommand{\TeV}{\ensuremath{\matqq	hrm{Te\kern -0.1em V}}}
\newcommand{\TeVc}{\ensuremath{\mathrm{Te\kern -0.1em V\!/}c}}
\newcommand{\TeVcc}{\ensuremath{\mathrm{Te\kern -0.1em V\!/}c^2}}
\newcommand{\GeV}{\ensuremath{\mathrm{Ge\kern -0.1em V}}}
\newcommand{\GeVc}{\ensuremath{\mathrm{Ge\kern -0.1em V\!/}c}}
\newcommand{\GeVcc}{\ensuremath{\mathrm{Ge\kern -0.1em V\!/}c^2}}
\newcommand{\MeV}{\ensuremath{\mathrm{Me\kern -0.1em V}}}
\newcommand{\MeVc}{\ensuremath{\mathrm{Me\kern -0.1em V\!/}c}}
\newcommand{\MeVcc}{\ensuremath{\mathrm{Me\kern -0.1em V\!/}c^2}}

\newcommand{\um}{\ensuremath{\mathrm{\mu m}}}

\newcommand{\cdfii}{CDF\,II~}

\newcommand{\myto}{\kern -0.3em\to\kern -0.2em}

\newcommand{\BR}{{\mathrm {BR}}}

\newcommand{\Bs}{\ensuremath{B^{0}_s}}

\newcommand{\phiphi}{\ensuremath{B_s^0 \myto \phi \phi}}
\newcommand{\psiphi}{\ensuremath{B_s^0 \myto J/\psi \phi}}
\newcommand{\psikst}{\ensuremath{B^0 \myto J/\psi K^{\ast 0}}}
\newcommand{\phikst}{\ensuremath{B^0 \myto \phi K^{\ast 0}}}
\newcommand{\kstar}{\ensuremath{K^{\ast 0} }}
\newcommand{\jpsi}{\ensuremath{J/\psi}}

\newcommand{\bspsikst}{\ensuremath{B_s^0 \myto J/\psi K^{\ast 0}}}
\newcommand{\bspsiks}{\ensuremath{B_s^0 \myto J/\psi K_S^0}}
\newcommand{\psiks}{\ensuremath{B^0 \myto J/\psi K_S^0}}

\newcommand{\DGs}{\ensuremath{\Delta \Gamma_s}}

\newcommand{\BVV}{\ensuremath{B\rightarrow VV\ }}
\newcommand{\BsVV}{\ensuremath{B_s^0\rightarrow VV}}

\newcommand{\fL}{\ensuremath{f_L\ }}

\title{Suppressed $B$ Decays at CDF}

\ShortTitle{HQL 2010}

\author{\speaker{Marco Rescigno}\\
        INFN, Rome \\
        E-mail: \email{marco.rescigno@roma1.infn.it}}

\abstract{
We present two recent results obtained by the CDF collaboration at
the Tevatron collider. New Cabibbo suppresed \Bs\ decay modes have been 
observed using 5.9 \ifb of data: \bspsiks\ and \bspsikst. 
We report also on masurement of the ratios of the branching ratios ($\BR$) of the new modes
to those of the $B^0$-meson to the same final states:
\begin{eqnarray*}
 \BR(\bspsikst)/\BR(\psikst) &=& 0.062 \pm 0.009 (stat.) \pm 0.025 (sys.) \pm 0.008 (frag.) 
\end{eqnarray*}
and
\begin{eqnarray*}
 \BR(\bspsiks)/\BR(\psiks)&=& 0.041 \pm 0.007 (stat.) \pm 0.004 (sys.) \pm 0.005 (frag.). 
\end{eqnarray*}
Then we discuss the first polarization measurement in a charmless
\Bs\ decay in two light vector mesons, \phiphi,
using 2.9 \ifb of data. An angular analysis of the final state particles
allows CDF to determine a longitudinal polarization fraction 
$\fL = 0.348 \pm 0.041 (stat.) \pm 0.021 (syst.)$, which is inconsistent 
with na\"{\i}ve expectations based on the V-A nature of weak currents and confirms 
the pattern of lower than expected
longitudinal polarization fraction in $b \myto s$ penguin dominated \BVV decays.
Finally, an updated measurement of the ratio of \phiphi\ $\BR$ to that of the
reference \psiphi\ mode is also presented:
\begin{eqnarray*}
 \BR (\phiphi)/\BR(\psiphi) = [1.78 \pm 0.14 (stat.) \pm 0.20 ( syst.)] \cdot 10 ^ {-2}.
\end{eqnarray*}
}

\FullConference{The Xth Nicola Cabibbo International Conference on Heavy Quarks and Leptons,\\
		October 11-15, 2010\\
		Frascati (Rome) Italy}

\begin{document}

\section{Introduction}

 The Tevatron collider has provided in the last decade an impressive amount of $p\bar{p}$ 
collision data that the two collaborations, CDF and D0, have very fruitfully  exploited. 
In particular large, samples of fully reconstructed \Bs\ decays have been collected 
allowing crucial progress on \Bs\ mixing, lifetime, decay width difference 
\DGs\ as well as the observation of a large number of decay modes. 
We will review here two recent results from the CDF experiment: 
the first observation 
of the Cabibbo suppressed \bspsiks\ and \bspsikst\ decay modes and measurement
of their branching ratio ($\BR$)~\cite{PN}, and the first angular analysis of 
charmless \phiphi\ decay for the determination of polarization amplitudes~\cite{phiphiPN}. 

 Important characteristics of the \cdfii\ detector~\cite{cdf} that are 
worth to be mentioned in connection to these two measurements are the trigger and 
charged track reconstruction capabilities.
A dimuon trigger with a $p_T$ threshold
as low as 1.5 \GeVc\ and $|\eta|<1$ is used for  $B\myto\jpsi X$ modes.
The trigger on displaced vertex with online measurement of
impact parameter of charged tracks~\cite{svt} allows the collection of hadronic decay 
modes like \phiphi.
The charged particles in the pseudorapidity 
range $|\eta|\lesssim 1$
are reconstructed by a silicon microstrip 
vertex detector and a drift chamber,
providing excellent resolution on $B$-meson decay length ($30\, \um$) and  
mass, typically about $10\,\, \MeVcc$ for $B\myto\jpsi X$ modes, 
that are crucial for the observation of rare \Bs\ modes.

\section{Observation of \bspsiks\ and \bspsikst}

The \psiks\ and \psikst\ decays are celebrated "golden" modes  where the greatly dominant 
amplitude is a Cabibbo favoured tree 
thus providing a crucial and theoretically clean determination of $\sin(2 \beta)$. Among the residual theoretical 
uncertainty, which may become important at the next generation flavor experiments, there are 
those related to the subleading 
penguin amplitude which is suppressed by $O(\lambda^2)$, where 
$\lambda = \sin(\theta_c)\sim 0.2$, with respect to the tree one. A different ratio between
tree and penguin is expected, on the other hand, in the 
\bspsiks\ and \bspsikst\ decays. For these modes the tree and penguin amplitudes enter both at order $O(\lambda)$. Consequently we 
expect a decay rate relative to the $B^0$ ones of order $O(\lambda^2)\sim 5-10\%$. Moreover,
by measuring both the rate and the CP violation in the \bspsiks\ mode, 
theoretical uncertainties on $\sin(2 \beta)$ due to penguin pollution will be 
reduced to a fraction of a 
degree~\cite{DeBruyn:2010hh}. Similar consideration apply to the
the study of \bspsikst\ to constrain theoretical uncertainties in the extraction of $\sin(2 \beta_s)$ from \psiphi\ decays~\cite{fleischer2}


The data used for this measurement corresponds to an integrated luminosity of 5.9~fb$^{-1}$. 
We derive the ratios of branching 
ratios of $\bspsiks$ and $\bspsikst$ to the reference $B^0$ decays using the relation:
\begin{equation*}
 \BR(\Bs \myto J/\psi K)/Br(B^0 \myto J/\psi K)=A_{rel}\times f_d/f_s\times N(\Bs \myto J/\psi K)/N(B^0\myto J/\psi K), 
\vspace{-0.1cm}
\end{equation*}
\noindent
where $K$ represents $K_S$ or $K^*$. By measuring the ratio of the number of $B^0$ and \Bs\ decays 
from data and the relative acceptance, $A_{rel}$, between the $B^0$ and \Bs\ modes from Monte Carlo simulation (MC), a
measurement of  $\BR(\Bs\myto \jpsi~K)/\BR(B^0\myto~\jpsi~K)$ is extracted using the ratio of fragmentation fractions $f_s/f_d$.
 

The event selection in the $B^0 \rightarrow J/\psi K^*$ analysis is optimized by maximizing $S/(1.5+~\sqrt B)$~\cite{punzi}
in a  simultaneous four-dimensional scan over four discriminating quantities:
$\pi$ $p_T$, $K$ $p_T$, transverse decay length $L_{xy}$ and $B$-vertex fit probability.
To extract the $B^0$ and \Bs\ signal yields 
a likelihood fit to the invariant mass distribution is performed.
The signal shape is modeled with three Gaussians template obtained from a fit to 
a simulated $B^0$ sample.
The \Bs\ template used in the fit is 
identical to $B^0$ template, except for a shift of 86.8 MeV/$c^2$ in the mean value of the three Gaussians,
corresponding to the mass difference between \Bs\ and $B^0$~\cite{pdg}.
The backgrounds considered in this analysis are combinatorial background, partially reconstructed background and \psiphi\ decay. 
The combinatorial background is modeled with an exponential function.
The partially reconstructed one, fitted with an ARGUS function~\cite{argus},  
is due to five-body decay with a $\pi$, $K$, or $\gamma$  not reconstructed. 
Finally,  a two Gaussians template, extracted from simulation, is used to 
model the \psiphi\ background, with a normalization constrained by data. 
The yields for $B^0$ and \Bs\ modes are respectively 9530 $\pm$ 110 
and 151 $\pm$ 25. The statistical significances of the \bspsikst\ signal is 8.0$\sigma$. 
The systematic uncertainty is dominated by the combinatorial background contribution, 
with a relative uncertainty 
for the $B^0$ to \Bs\ ratio of 31.4\%. Other sources of systematic uncertainty are the signal modeling (4.4\%), and $\psiphi$ contribution (1.3\%). 

\begin{figure} 
\center
\includegraphics[width=0.42\linewidth]{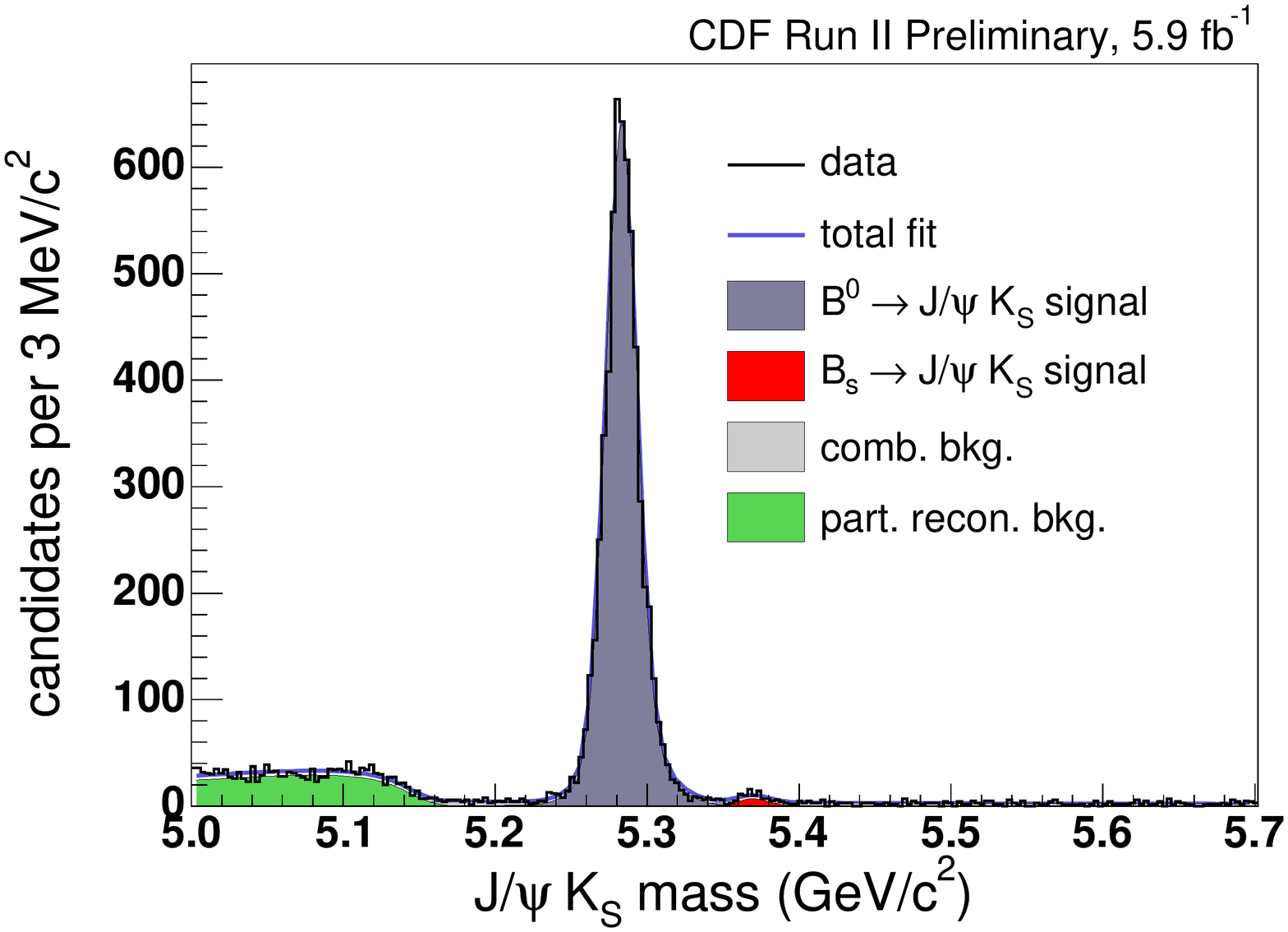} 
\includegraphics[width=0.42\linewidth]{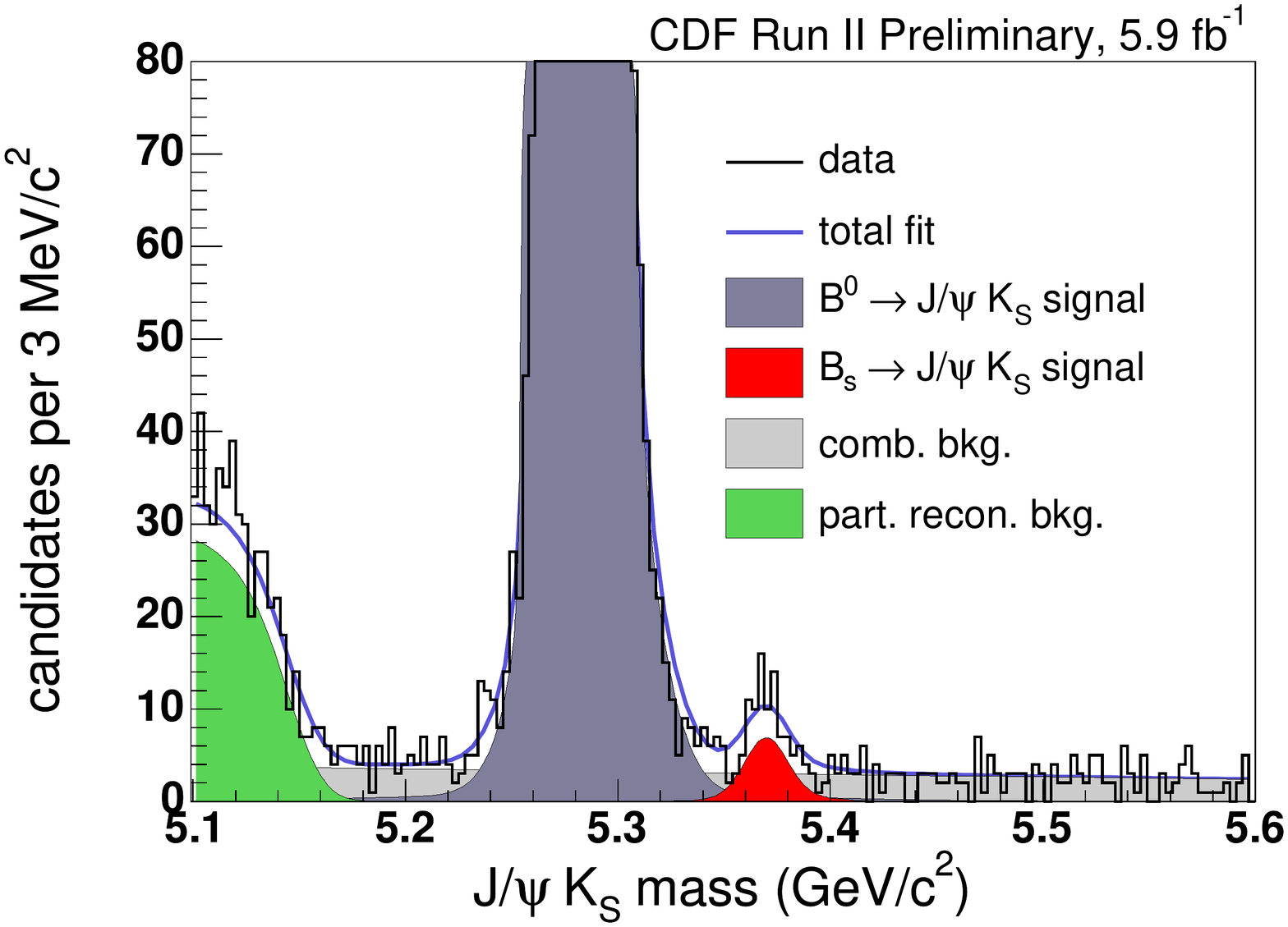} 
\caption{\label{fig1}
Invariant mass for selected $B \myto \jpsi K^0_S$ candidates with fit (left). Enlarged version of the same plot showing
the \Bs\ signal region in greater detail (right). 
}
\end{figure}

For the observation of the \bspsiks\  a Neural Network (NN) based multivariate classifier 
is used to further reduce combinatorial background.
%
%
%
In order to train the NN, simulated \Bs\ MC events are used as signal. 
Data from the upper side band in the \Bs\ candidate invariant mass distribution, well separated from the signal region, 
are used as a background data sample. 
A likelihood fit similar to the one described before is performed to the invariant mass distribution to extract the yield of the \psiks\ and \bspsiks\ signals. 
From the fit, shown in Fig.~\ref{fig1}, the yields of the $B^0$ and 
\Bs\ signal are determined to be 5954 $\pm$ 79 and 64 $\pm$ 14, respectively.
The statistical significances of the \bspsiks\ signal is 7.2$\sigma$. 
In this case the relative uncertainties for the ratio of yields are 5.6\% from the 
combinatorial background contribution, 5.6\% from the combinatorial background modeling, 
4.6\% from the signal modeling.

To determine the $\BR(\Bs \myto J/\psi K)/Br(B^0 \myto J/\psi K)$, 
a correction of 1\% and 5\% for the ratio of 
acceptance is obtained from simulation respectively for the $J/\psi K^0_S$
and $J/\psi K^{0\ast}$ case.
The \Bs\ and $B^0$ lifetimes, B hadron $p_T$ spectrum and polarization, this last one only for the \bspsikst\
analysis, are considered as a source of systematic uncertainty in the efficiency correction.
The most recent CDF measurement~\cite{fracCDF} of $f_s/(f_u+f_d)\times Br(D_s \rightarrow \phi \pi )$ is combined with the 
actual PDG value~\cite{pdg} for $Br(D_s \rightarrow \phi \pi )$, to extract  $f_s/f_d$ = 0.269 $\pm$ 0.033. We can
thus estimate:
\vspace{-0.2cm}
\begin{eqnarray*}
 \BR(\bspsikst)/\BR(\psikst) &=& 0.062 \pm 0.009 (stat.) \pm 0.025 (sys.) \pm 0.008 (frag.) \\
\vspace{-0.6cm}
 \BR(\bspsiks)/\BR(\psiks)&=& 0.041 \pm 0.007 (stat.) \pm 0.004 (sys.) \pm 0.005 (frag.). 
\end{eqnarray*}
This confirms the order of magnitude estimate, $O(\lambda^2$), for this ratio given above. 


\section{\phiphi\ Polarization Measurement}
 The $\phiphi$ belongs to a particular class, \BsVV, of decays  in a 
 pair of J=1 mesons which are in a superposition of CP eigenstates.
 It will be used to constrain new physics contribution to  
 \Bs\ mixing phase through a measurement of time dependent
 CP violation. 
 Three independent amplitudes govern \BsVV\ decays, 
 corresponding to the possible polarizations of the 
 final state mesons. It is thus attractive to test the 
 theoretical predictions for these polarization 
 amplitudes~\cite{Beneke:2006hg,Ali:2007ff,Cheng:2009mu}.
%
  Evidence for the $\phiphi$ process has been reported by CDF with low statistics~\cite{prlphix}.
 We discuss here the first measurement of polarization amplitudes in this decay
 and an updated measurement of the branching ratio using 2.9 \ifb.
  The $\phiphi$ decay proceeds through a $b\rightarrow s\overline{s}s$ quark level
 process, and, in the Standard Model (SM), the dominant diagram is the $b\rightarrow s$
 penguin. The same penguin amplitude is involved in several processes
 which have shown several discrepancies with the SM predictions. 
 In particular, both  SM and new physics interpretations
 have been considered to explain the  lack of dominant longitudinal polarization 
 component for several penguin dominated \BVV\ decay modes~\cite{puzzle}. 
 Measurements of polarization amplitudes  in new modes, including \phiphi\ decays,
 have been proposed~\cite{Datta:2008wf} to resolve this issue.
 We study also \psiphi\ decays in the same dataset, use this mode as a normalization 
 for the \phiphi\ BR measurement and  extract \psiphi\ polarization amplitudes
 as a cross check of the main \phiphi\ result.

Event selection is the same for both BR and polarization measurement and is described
in detail elsewhere~\cite{PNphiphiBR}.
The invariant mass of the selected \phiphi\ candidates is shown in 
Fig.~\ref{fig:projection} along with the projection of the likelihood fit 
desribed in the following.
Two sources of background are expected in the  $B_s^0$~signal region: 
combinatorial background 
and \phikst\  reflection with the wrong assignment of a kaon mass to the \kstar\ decay pion. Similar
consideration apply for the \psiphi\ case where  \psikst\ constitute the only
expected reflection component.
%
We estimate a contribution of $f_{\psikst}=(4.19 \pm 0.93) \%$
and $f_{\phikst}=(2.7 \pm 1.0) \%$ under respectively the \psiphi\
and \phiphi\ signals and fit the total number of signal decays as 
$N_{\phi \phi} = 295 \pm 20 \pm 12$  
and $N_{\psi \phi} = 1766 \pm 48 \pm 41$ where the first uncertainty is 
statistical and the second
is systematic and is evaluated using alternative signal and background models.\newline
%
\indent
 To extract the \phiphi\ decay rate first the measurement of the branching ratio 
ratio to the \psiphi\ mode is performed by correcting for the relative detection
efficiency for the two decays:
%
 $ \BR (\phiphi)/\BR(\psiphi) = [1.78 \pm 0.14 ({\rm stat.}) \pm 0.20 ({\rm syst}.)] \cdot 10 ^ {-2} $.
We then derive $ \BR (\phiphi) = [2.40 \pm 0.21 ({\rm stat.}) \pm 0.27 ({\rm syst.}) \pm 0.82 (\BR)] \cdot 10 ^{-5} $,
adopting the BR(\psiphi) from ref.~\cite{CDF-psiphiBR}, 
corrected for the current measurements~\cite{pdg} of 
$f_s/f_d$\footnote{We actually use: $\BR(\psiphi)=(1.35 \pm 0.46)\cdot 10^{-3}$}.
The dominant systematic uncertainty, labeled (BR), originate from the 
$\BR(\psiphi)$ uncertainty alone. 
This result is in agreement and supersedes our previous measurement\cite{prlphix} 
and represents a substantial improvement in the statistical uncertainty;
it is as well compatible with recent theoretical predictions~\cite{Beneke:2006hg,Ali:2007ff,Cheng:2009mu}.
%
%
%
\begin{figure} [t]
\begin{center}
\includegraphics[width=0.4\linewidth]{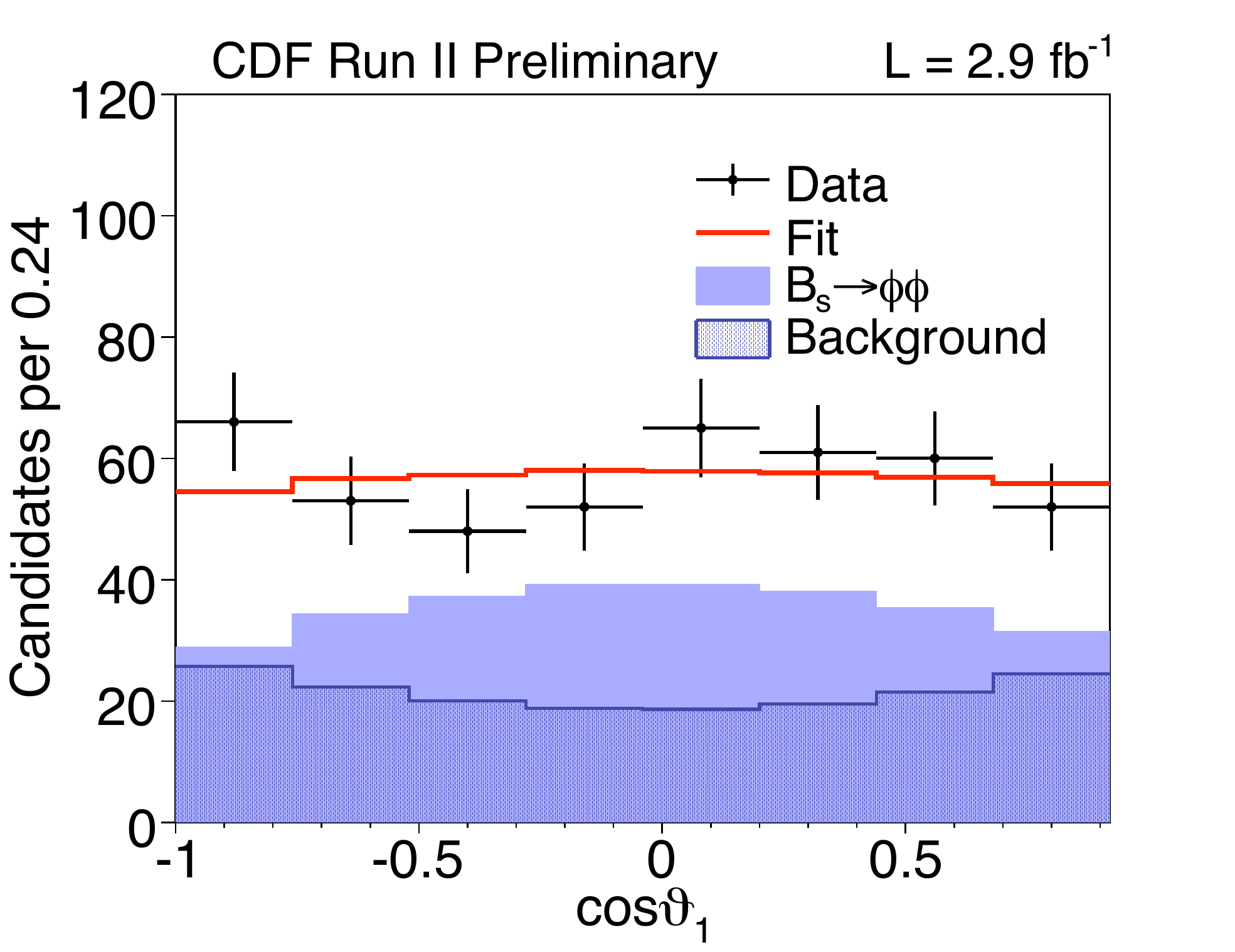} 
\includegraphics[width=0.4\linewidth]{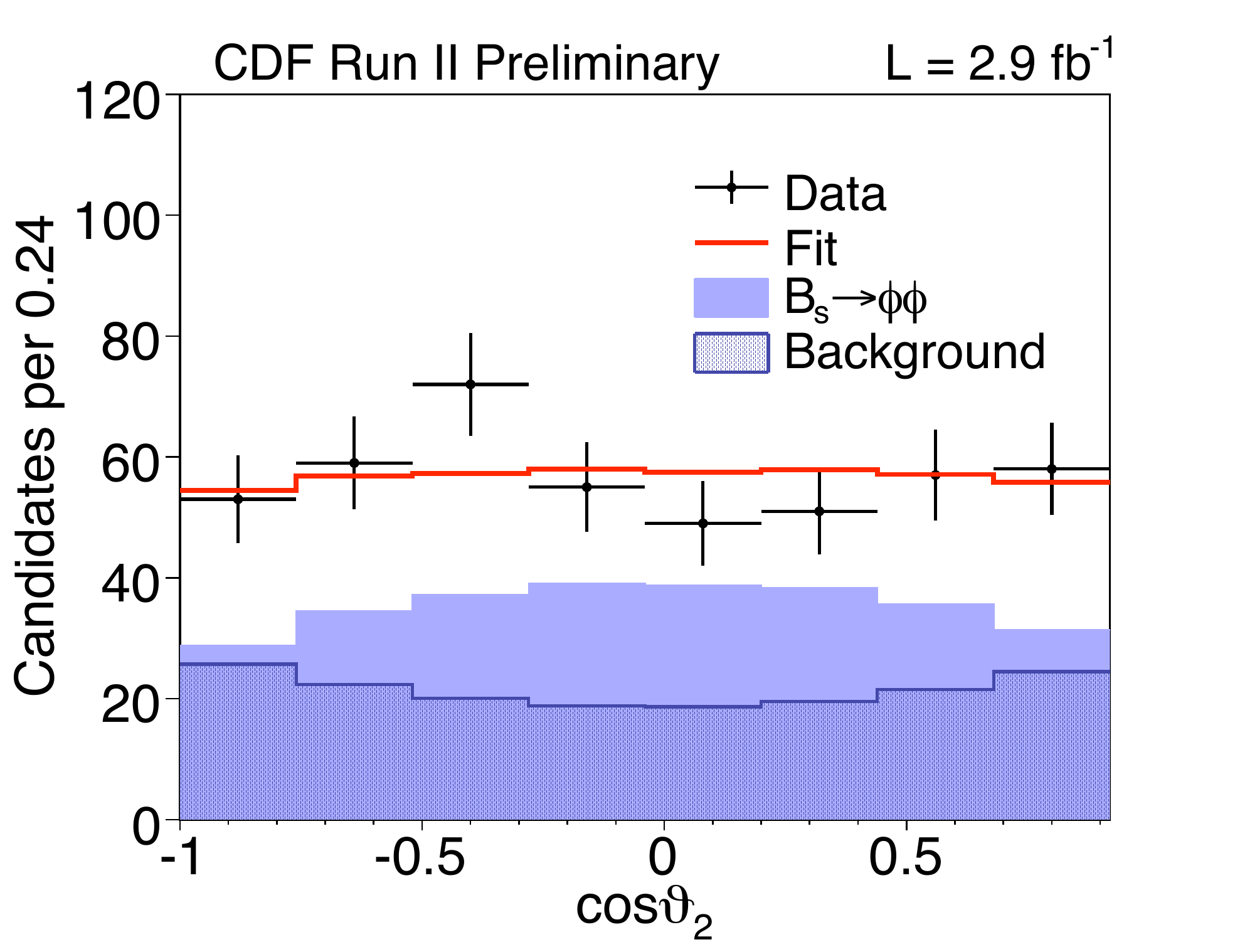} 
\includegraphics[width=0.4\linewidth]{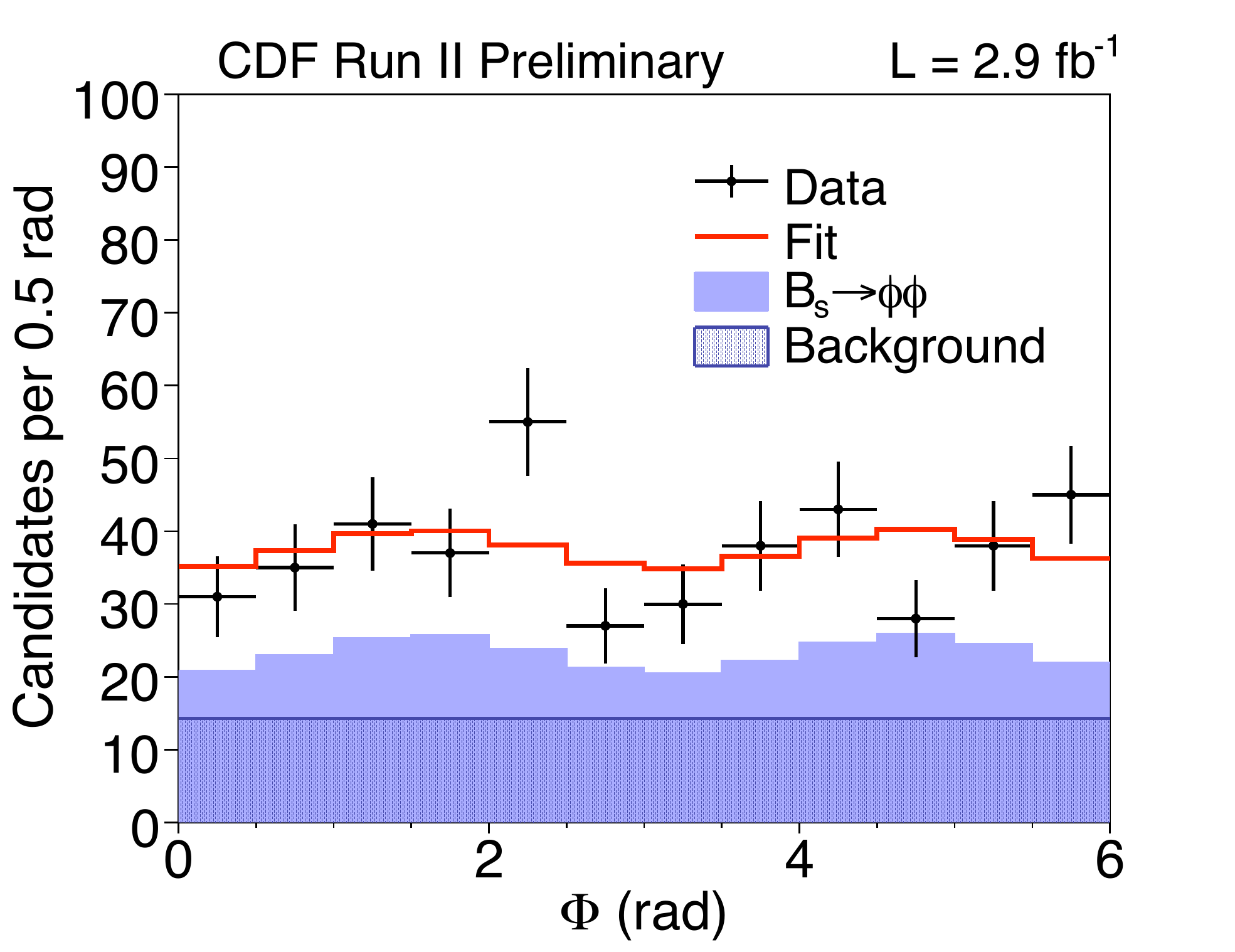} 
\includegraphics[width=0.4\linewidth]{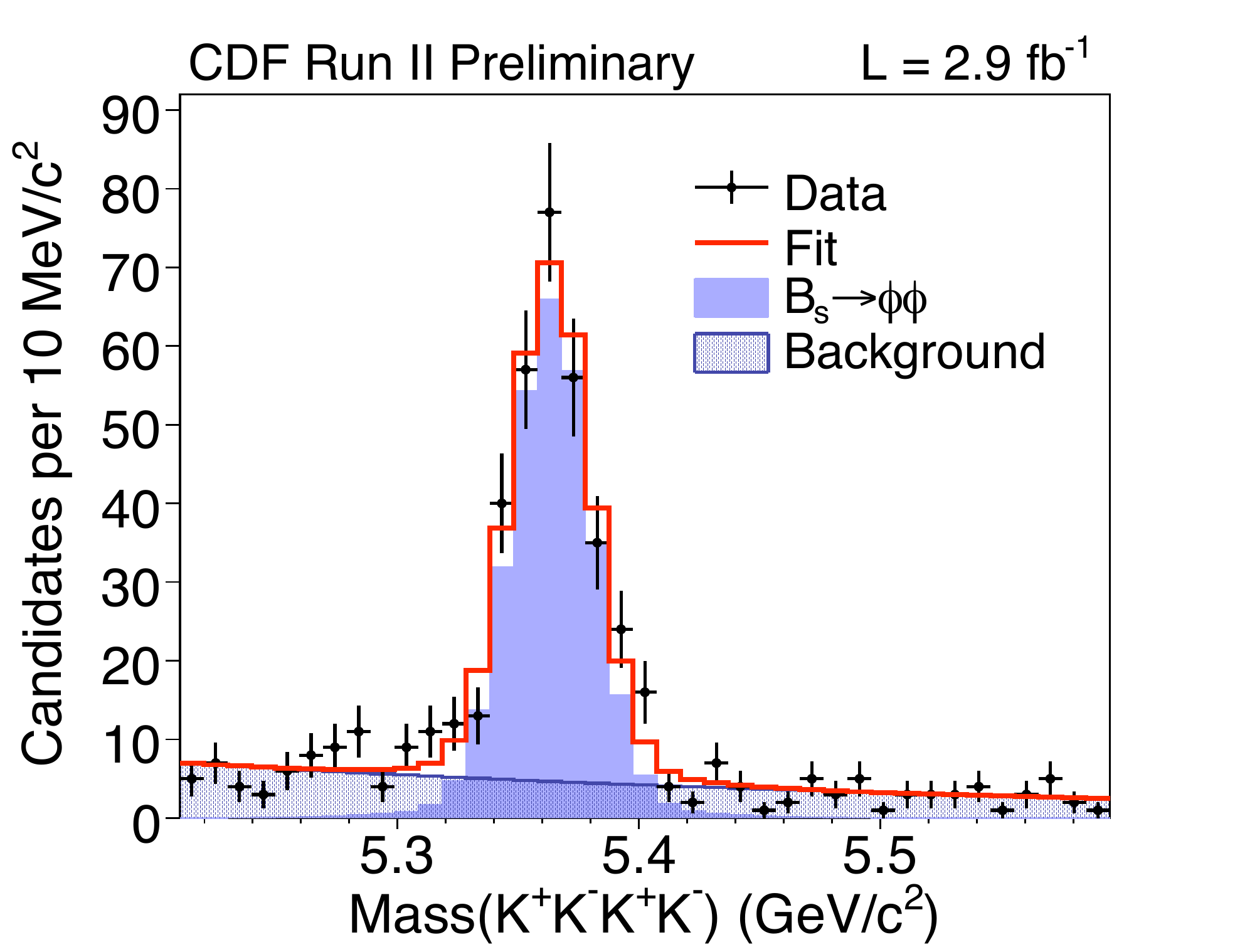} 
\caption{\label{fig:projection}
\phiphi\ candidate mass and angular distribution with overlayed fit result 
for signal and background.}
\end{center}
\vspace{-0.8cm}
\end{figure}

  The angular distribution of the \phiphi\ decay products can be
described using the helicity variables,
$\vec{\omega}=(\cos\vartheta_1,\cos\vartheta_2,\Phi$), where
$\vartheta_i$ is the angle between the direction of the $K^+$ from each
$\phi \to K^+K^-$ and the direction opposite the $B^0_s$ in the vector
meson rest frame, while $\Phi$ is the angle between the two resonance
decay planes. The total decay amplitude can be decomposed in three complex amplitudes
$H_\lambda$ corresponding to the vector helicity $\lambda=0,\pm1$; 
we use their linear combinations which give the polarization 
amplitudes\footnote{The polarization amplitudes are normalized
so that the following condition holds:
$|A_0|^2+|A_\parallel|^2+|A_\perp|^2=1$.}:
$A_0=H_0$, $A_\parallel=(H_++H_-)/\sqrt{2}$ and
$A_\perp=(H_+-H_-)/\sqrt{2}$. The differential decay rate can
be expressed as 
$\frac{d^4\Gamma}{dtd\vec{\omega}} \propto \sum_{i=1}^6 K_i(t)f_i(\vec{\omega})$
where the $K_i(t)$ terms account for the exponential decay and the
time evolution of the \Bs\ state due to mixing and decay width differences \DGs\
while the $f_i(\vec{\omega})$ are functions of the helicity
angles only. We measure the untagged decay rate integrated in time and
neglect the \Bs\ mixing phase (tiny in the SM) and assume no direct CP violation. 
The differential decay rate then  
depends on the polarization amplitudes at $t=0$ and on
the light and heavy \Bs\ mass-eigenstate lifetimes, $\tau_{\text{L}}$ and $\tau_{\text{H}}$
respectively, as follows: 
\begin{equation}
\begin{split}
\label{eq:decay_rate}
\frac{d^3\Gamma}{d\vec{\omega}} \propto &\, \tau _{\textup{L}}\big(
|A_0|^2f_1 (\vec{\omega}) + |A_\parallel|^2 f_2 (\vec{\omega}) +\\ 
& + |A_0||A_\parallel|\cos\delta_\parallel f_5 (\vec{\omega})\big) + \tau_{\textup{H}}|A_\perp|^2f_3 (\vec{\omega}),
\end{split}
\end{equation}
where $\delta_\parallel=\arg (A_0^\star A_\parallel)$.
 and
\begin{eqnarray*}
\begin{split}
f_1(\vec{\omega})&=4\cos^2\vartheta_1\cos^2\vartheta_2, 
& f_2(\vec{\omega})&=\sin^2\vartheta_1\sin^2\vartheta_2(1+\cos 2\Phi),\\
f_3(\vec{\omega})&=\sin^2\vartheta_1\sin^2\vartheta_2(1-\cos 2\Phi),
& f_5(\vec{\omega})&=\sqrt{2}\sin 2\vartheta_1 \sin 2\vartheta_2 \cos\Phi.
\end{split}
\end{eqnarray*}
%
%
%

We perform an unbinned maximum likelihood fit to the reconstructed
mass $m$ of the \Bs\ candidates and the helicity
variables in order to measure the polarization amplitudes.
The mass distribution is used in the fit to discriminate the
signal from the background. The identification of the two $\phi$ as
$\phi_1$ and $\phi_2$ to define the angles $\vartheta_1$ and
$\vartheta_2$ is randomly implemented in order to satisfy the Bose
symmetry under indexes exchange $1 \leftrightarrow 2$. The likelihood
for each candidate is defined as
$\mathcal{L}_i=(1-f_b)\mathcal{P}_s(m_i,\vec{\omega}_i|\vec{\xi}_s)+f_b\mathcal{P}_b(m_i,\vec{\omega}_i|\vec{\xi}_b)$,
where $f_b$ is the fraction of the background and
$\mathcal{P}_j(m_i,\vec{\omega}_i|\vec{\xi}_j)$ are the probability density function
(PDF) for the \phiphi\ signal ($j=s$) and background ($j=b$) components which
depend on the fitting parameters, $\vec{\xi}_s$ and $\vec{\xi}_b$
respectively. 
Both the signal and the background PDF are the product of a mass PDF and an
angular one. 
The signal mass component for both signal and background has been 
already described.
Fixing $\tau_{\textup{L}}$ and $\tau_{\textup{H}}$ to the world
average values~\cite{pdg}, the angular
PDF for the signal is given by Eq.~\ref{eq:decay_rate} multiplied by
an acceptance factor; the latter is implemented as a three-dimensional
histogram representing the probability to find an event at each
position of the $\vec{\omega}$ space. The angular acceptance is derived from a MC
simulation of the \phiphi\ decay, generated averaging over all possible spin states
of the decay products.
We use a purely
empirical parameterization derived by analysing the angular
distributions in the mass sidebands to model the background angular
PDF.
%
%
The fitter is extensively tested using simulated
samples with a variety of input parameters.
A further check is performed 
by repeating the same measurement for the \psiphi\ events 
collected with the same displaced vertex trigger as \phiphi;
we find $|A_0|^2=0.534\pm0.019\text{(stat)}$ and $|A_\parallel|^2=0.220\pm0.025\text{(stat)}$, 
in very good agreement with CDF and D0 measurements~\cite{psiphiparam}. 
%
\begin{figure} 
\begin{center}
\begin{tabular}{lc}
     Observable        &  Result \\
\hline
$\BR$                  &  $[2.40 \pm 0.21  \pm 0.85 ] \cdot 10 ^{-5} $ \\
$|A_0|^2$              & $0.348 \pm 0.041 \pm 0.021$ \\
$|A_\parallel|^2$      & $0.287 \pm 0.043 \pm 0.011$ \\
 $|A_\perp|^2$ & $0.365 \pm 0.044 \pm  0.027$   \\
$\cos\delta_\parallel$ & $-0.91^{+0.15}_{-0.13} \pm 0.09$    \\
\hline
\vspace{5.9cm}
\end{tabular}
\includegraphics[width=0.55\linewidth]{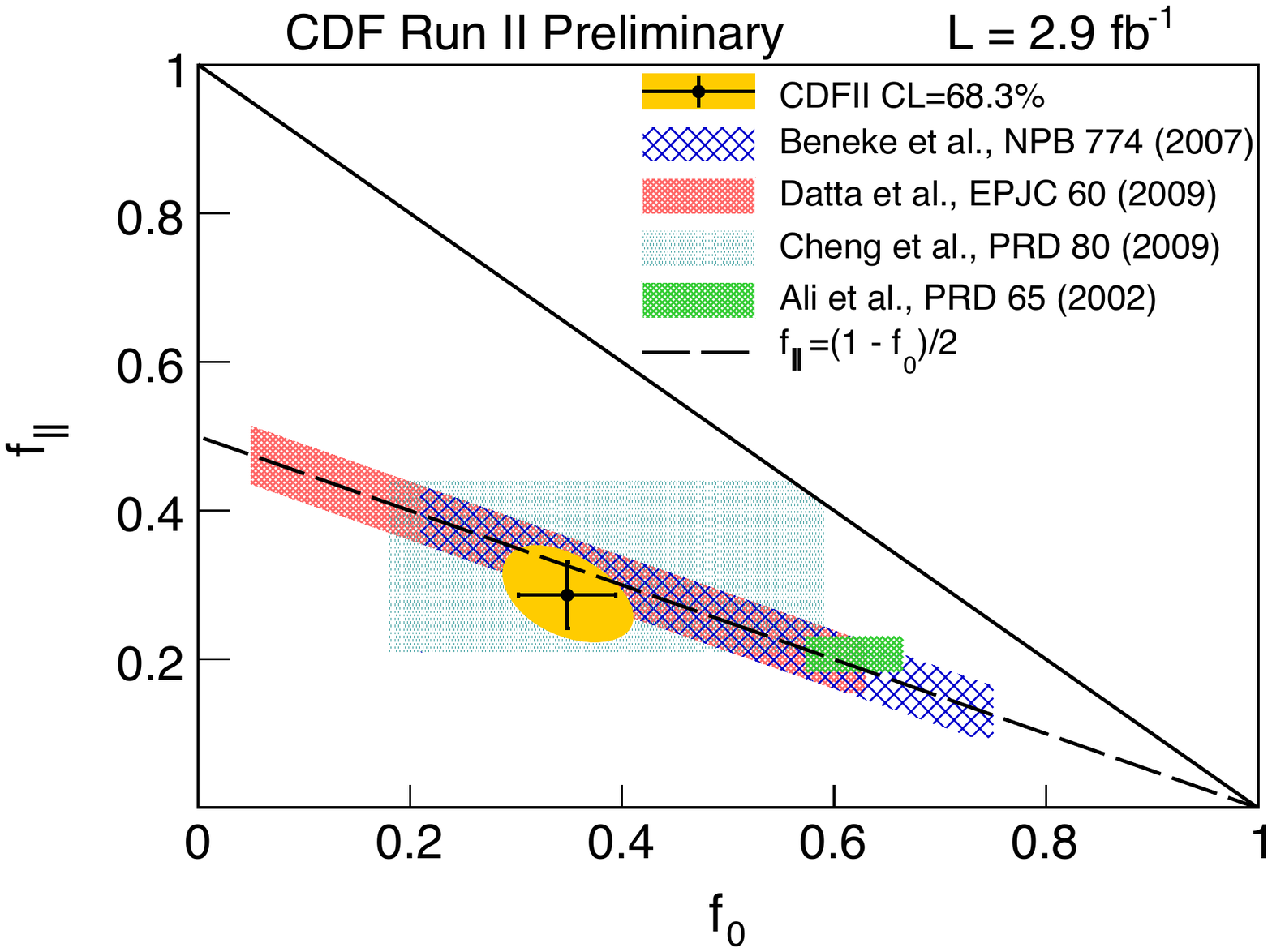} 
\end{center}
\vspace{-5.0cm}
\caption{\label{fig:fittheory}
\phiphi\ experimental results with stat. and syst. uncertainties (left panel), 
comparison to recent theory predictions (right panel).}
\end{figure}

 Fit projections on the angular variables and the results for the 
polarization observables compared to recent theory
calculations are shown in  Fig.~\ref{fig:projection} and Fig.~\ref{fig:fittheory}.
%
%
Several sources of systematic uncertainty are considered.
We account for the physics background effects through simulated
samples. We consider the \phikst\ decay, 
the resonant $B^0_s \to \phi f_0(980)$ decay and the decay 
$B^0_s$ to $\phi$ plus a non-resonant kaon pair,
whose fractions are normalized to the signal yield in analogy with similar $B^0$ decays.
Assuming up to  4.6\% contamination of $B^0_s\to\phi f_0$ and 
0.9\% of $B_s^0\to \phi(K^+K^-)$ we estimate a 1.5\% 
systematic uncertainty from backgrounds unaccounted for. 
Possible biases introduced by the time integration are
examined with MC simulation: they are induced by the dependence of the angular acceptance on
$\Delta\Gamma_s$ and by a non-uniform acceptance with the $B^0_s$
proper decay time introduced by the displaced track trigger; the assigned
systematic (1\%) is the full shift expected in central value,
assuming a value for $\Delta\Gamma_s$ equal to the world average plus
one standard deviation. We consider the propagation of $\tau_{\textup{L(H)}}$
uncertainties to the polarization amplitudes (1\%). We have also 
verified that the impact of a sizeable CP-violating phase in \Bs\ mixing
would be negligeable on polarization observables. 
Combinatorial background parametrization and angular acceptance
give minor contributions to systematic uncertainties.
\newline
\indent 
In conclusion for \phiphi\ we find a significantly suppressed longitudinal fraction
$f_{\text{L}}=|A_0|^2 =0.348\pm0.041\text{(stat)}\pm0.021\text{(syst)}$, that is 
found to be even smaller than in other $b \myto s$ penguin \BVV decays~\cite{pdg}. 
This result is 
in agreement, within uncertainties, with predictions~\cite{Beneke:2006hg,Cheng:2009mu} 
based on QCD factorization, but seems to contradict others~\cite{Ali:2007ff}. 
It implies the hyerarchy $H_0 \simeq H_+ >> H_-$ in polarization amplitudes,
possibly induced by a large penguin annihilation contribution~\cite{puzzle,Datta:2008wf}.

\end{document}